\documentclass[letterpaper, fontsize=12pt]{scrartcl}	
\usepackage[margin=1.00in]{geometry}
\usepackage[T1]{fontenc}
\usepackage{fourier}
\usepackage[english]{babel}															
\usepackage[protrusion=true,expansion=true]{microtype}				
\usepackage{amsmath,amsfonts,amsthm,amssymb}	
\usepackage[flushleft]{threeparttable}
\usepackage{longtable}
\usepackage{pdflscape}
\usepackage{multicol}									
\usepackage[pdftex]{graphicx}														
\usepackage{url}
\usepackage{wrapfig}
\usepackage{subcaption}
\usepackage{booktabs}
\usepackage{siunitx}
\usepackage{enumitem}
\setlist[enumerate]{itemsep=0mm}
\setlist[itemize]{itemsep=0mm}
\usepackage[normalem]{ulem}
\usepackage{pdfpages}
\usepackage{fancyhdr}
\usepackage[numbers,sort&compress]{natbib}
\usepackage{twoopt}

\usepackage{titlesec, blindtext, color}
	\definecolor{light-blue}{rgb}{0.8,0.85,1}
	\definecolor{airforceblue}{rgb}{0.36, 0.54, 0.66}
	\definecolor{azure}{rgb}{0.0, 0.5, 1.0}
	\definecolor{bleudefrance}{rgb}{0.19, 0.55, 0.91}
	\definecolor{blue(munsell)}{rgb}{0.0, 0.5, 0.69}
	\definecolor{darkmidnightblue}{rgb}{0.0, 0.2, 0.4}
	\definecolor{steelblue}{rgb}{0.27, 0.51, 0.71}
	\definecolor{tealblue}{rgb}{0.21, 0.46, 0.53}
	\definecolor{yaleblue}{rgb}{0.06, 0.3, 0.57}
	\definecolor{applered}{rgb}{0.89, 0.02, 0.17}

\usepackage{sectsty}												
\allsectionsfont{\normalfont \large \scshape \color{yaleblue} \bfseries}	
\usepackage{setspace}

\usepackage[pdftex,unicode, implicit]{hyperref}
\hypersetup{
    unicode=true,          
    linktocpage=true,
    pdftoolbar=true,        
    pdfmenubar=true,        
    pdffitwindow=true,     
    pdfstartview={FitH},    
    pdfnewwindow=true,      
    colorlinks=true,       
    linkcolor=yaleblue,          
    citecolor=yaleblue,        
    filecolor=yaleblue,      
    urlcolor=yaleblue, 
   dvipdfm=true
}

\newcommand{\horrule}[1]{\rule{\linewidth}{#1}} 	

\title{         \normalfont 							
		\vspace{-1.25in} 	
		{\color{yaleblue}\horrule{2pt}} \\
		\LARGE	
		{\color{yaleblue}{\textsc{{\textbf{Next-Generation Comprehensive Data-Driven Models of Solar Eruptive Events}}}}}\\ 
	
                \vspace{.1in}
        		\small
		A white paper in response to the \textbf{Solar and Space Physics (Heliophysics) Decadal Survey}\\
		\color{applered}{SUBMITTED. No further changes can be made.}
		\vspace{-0.05in}
		{\color{yaleblue}\horrule{2pt}}
		\large
		\color{black}
		\textbf{Principal Author}\\
                Joel C. Allred$^{1}$ (joel.c.allred@nasa.gov)\\
                \vspace{.1in}
                \textbf{Co-authors}\\
                Graham S. Kerr$^{1,2}$, Meriem Alaoui$^{1,3}$, Juan Camilo	Buitrago-Casas$^{4}$, Amir Caspi$^{5}$, Bin Chen$^{6}$,
                Thomas Y. Chen$^{7}$, Lindsay	Glesener$^{8}$, Silvina E. 	Guidoni$^{9}$, Fan Guo$^{10}$, Judith T. Karpen$^{1}$, Sophie Musset$^{11}$, Katharine K. Reeves$^{12}$, Albert Y. Shih$^{1}$\\
                \vspace{.05in}
                \small
                \raggedright
                \textsl{(1) NASA/GSFC, (2) Catholic University of America, (3) University of Maryland, (4) University of California Berkeley, (5) Southwest Research Institute, (6) New Jersey Institute of Technology, (7) Columbia University, (8) University of Minnesota, (9) American University, (10) Las Alamos National Lab, (11) European Space Agency ESTEC, (12) Center for Astrophysics Harvard \& Smithsonian}
}

\newcommand{\ion}[2]{#1~\textsc{#2}}

\newcommand{\bluefont}{
   \color{yaleblue}
}
\DeclareTextFontCommand{\blue}{\bluefont}

\newcommand{\kglobal}{\emph{kglobal{}}}
\newcommand{\arms}{\emph{ARMS}}
\newcommand{\radyn}{\emph{RADYN}}
\newcommand{\fp}{\emph{FP}}

\newcommand{\radynarcade}{\emph{RADYN\_Arcade}}

\newcommand{\flarix}{\emph{FLARIX}}
\newcommand{\hydrad}{\emph{HYDRAD}}
\newcommand{\muram}{\emph{MURaM}}
\date{}

\begin{document}
\begin{spacing}{1}
\maketitle
\vspace{-1in}
\end{spacing}
\thispagestyle{empty}
\section*{Abstract}
\vspace{-.15in}
Solar flares and coronal mass ejections are interrelated phenomena that together are known as solar eruptive events. These are the main drivers of space weather and understanding their origins is a primary goal of Heliophysics. In this white paper, we advocate for the allocation of sufficient resources to bring together experts in observations and modeling to construct and test next generation data-driven models of solar eruptive events. We identify the key components necessary for constructing comprehensive end-to-end models including global scale 3D MHD resolving magnetic field evolution and reconnection, small scale simulations of particle acceleration in reconnection exhausts, kinetic scale transport of flare-accelerated particles into the lower solar atmosphere, and the radiative and hydrodynamics responses of the solar atmosphere to flare heating. Using this modeling framework, long-standing questions regarding how solar eruptive events release energy, accelerate particles, and heat plasma can be explored.

\textbf{To address open questions in solar flare physics, we recommend that NASA and NSF provide sufficient research and analysis funds to bring together a large body of researchers and numerical tools to tackle the end-to-end modeling framework that we outline. Current dedicated theory and modeling funding programs are relatively small scale and infrequent; funding agencies must recognize that modern space physics demands the use of both observations and modeling to make rapid progress.} 
\newpage
\setcounter{page}{1}

\vspace{-0.15in}
\section{Introduction}
\vspace{-0.15in}
Solar flares and coronal mass ejections (CMEs) are the major drivers of space weather, sharing a common energy source (the coronal magnetic field), a common energy-release mechanism (magnetic reconnection), and comparable total energies \citep{2012ApJ...759...71E}. Predicting their development, and ultimately their contribution to space weather, requires a unified and comprehensive understanding of both flares and CMEs as interrelated phenomena that together are referred to as Solar Eruptive Events (SEEs).

Understanding SEEs is not only a core goal of Heliophysics but is critical for making reliable space-weather predictions needed for future space exploration, given SEEs' potential to expose astronauts to harmful radiation and damage spacecraft. During the past solar cycle, NASA's Heliophysics System Observatory has provided unprecedented views of the critical processes at play in SEEs. Parallel developments in theory and computing resources have greatly advanced our ability to both model these processes and predict observables. These capabilities have now reached a stage sufficient to develop a data-driven model from SEE initiation through the flare gradual phase. Moreover, improved data-analysis techniques are enabling the extraction of more information from the observations than was previously possible. 

Given these recent advances, in this white paper, we advocate for the development of comprehensive data-driven coupled models for performing holistic studies of SEEs. These will provide a more complete picture and understanding of energy release and partition in SEEs, especially on the fundamental energy-release processes associated with the magnetic reconnection that leads to the flare and the initial acceleration of the CME. \textbf{To enable this, we recommend that NASA and NSF provide sufficient research and analysis funds to bring together a large body of researchers and numerical tools to tackle this problem. Current dedicated theory and modeling programs are relatively small scale and infrequent; funding agencies must recognise that modern space physics demands the use of both observations and modeling to make rapid progress.} 

\vspace{-0.175in}
\section{Current Generation SEE Models}
\vspace{-0.15in}
Global-scale 3D models of SEE initiation invoke magnetic reconnection in the flare current sheet (CS), which restructures the magnetic field adjacent to the CS, transforming stored magnetic energy into mass motions, particle acceleration, and heating. High-resolution observations and numerical simulations are increasingly able to determine the mechanics of this process, but we still do not fully understand how the liberated energy powers the many manifestations of the flare and CME. An important clue derived from observations and MHD modeling is the appearance of many reconnection sites within the flare CS, yielding multiple plasmoids (magnetic islands) that travel upward or downward in the sheet, within the fast bidirectional reconnection exhausts \citep{2012ApJ...760...81K, 2013A&A...557A.115K, 2016ApJ...820...60G, 2016ApJ...826...43L}. Fermi acceleration resulting from contracting and merging plasmoids has been shown to produce power-law distributions similar to those observed \citep{2021PhRvL.126m5101A}. Alternatively, stochastic acceleration from magnetic turbulence also produces power-law distributions \citep{2012SSRv..173..535P}. It is not yet known, which or if either, of these processes dominate flare particle acceleration.

During the flare reconnection process, electrons and ions are accelerated. The energetic electrons are detected through the gyrosynchrotron microwave (MW) emission and the hard X-ray (HXR) bremsstrahlung that are produced as they propagate along magnetic field lines from the coronal acceleration site to their eventual impact in the chromospheric footpoints. The energy distribution of the impacting electrons can be readily determined from the X-ray spectrum \citep{2011SSRv..159..107H}. This energy distribution will differ from that of the electrons just leaving the acceleration region, due to energy losses from Coulomb collisions, return currents, and other processes as the particles travel from the acceleration site. Models that account for all of these transport effects are essential for any meaningful comparison of electron distributions inferred from the microwave and X-ray observations with predictions from acceleration models.

The accelerated ions are detected through the nuclear $\gamma$-ray emission that is produced as they interact with the ambient plasma. Observations suggest that these ions have approximately the same total energy content as the accelerated electrons, at least in large flares \citep{2012ApJ...759...71E,2017ApJ...836...17A}. However, the $\gamma$-ray emission from the ions is not as easily detectable as the microwave and X-ray emission from the electrons, making it much more challenging to measure their energy distributions, and virtually impossible below $\sim$1 MeV. Consequently, very few theoretical studies have modeled flare-accelerated ions or their effects on the solar atmosphere. However, for an acceleration mechanism to be viable, it must be able to predict both the ion and electron distributions. Previous work neglecting ions is potentially ignoring up to half of the flare energy transported through the Sun's atmosphere \citep{2021PhRvL.127r5101Z}. Next generation models must explore the role of ion beam heating in flares, with a specific focus on the resulting heating of the deepest layers of the solar atmosphere (see also the white paper, ``Ion Acceleration in Solar Eruptive Events'' by Shih et al.).

SEEs heat plasma to super-hot temperatures \citep{2014ApJ...781...43C} ($\gtrapprox 30$~MK). This super-hot plasma appears to be directly heated in the corona \citep{2010ApJ...718.1476G, 2011ApJ...740...73L}, but the responsible mechanism has not been established. This heating mechanism likely works in concert with the particle acceleration mechanism mentioned above, and both are signatures of the fundamental energy release driving SEEs. Next generation models should investigate how SEEs directly heat plasma during the acceleration and energy release processes.

\vspace{-0.175in}
\section{Open Science Questions}
\vspace{-0.15in}
Comprehensive SEE models can probe the fundamental energy release processes by tracking the released energy in the form of energetic particles and thermal plasma as they manifest themselves in the flare and CME. Using these models the following long-standing questions can be explored. These get to the heart of how energy is released, particles are
accelerated, plasma is heated, and mass motions are driven in magnetized plasmas on the Sun.

\vspace{0.1in}
\noindent 1. \textbf{What is the relation between magnetic reconnection, magnetic energy release, and SEE eruption and acceleration?}

\vspace{0.1in}
\noindent 2. \textbf{How are particles accelerated during flares? What are the energy distributions of accelerated electrons and ions and how do they evolve in time and space?}

\vspace{0.1in}
\noindent 3. \textbf{How are flare plasmas heated? How is direct in-situ heating related to particle acceleration?}
\vspace{0.1in}

\newpage
\begin{landscape}
\begin{figure}[!ht]
\includegraphics[scale=1]{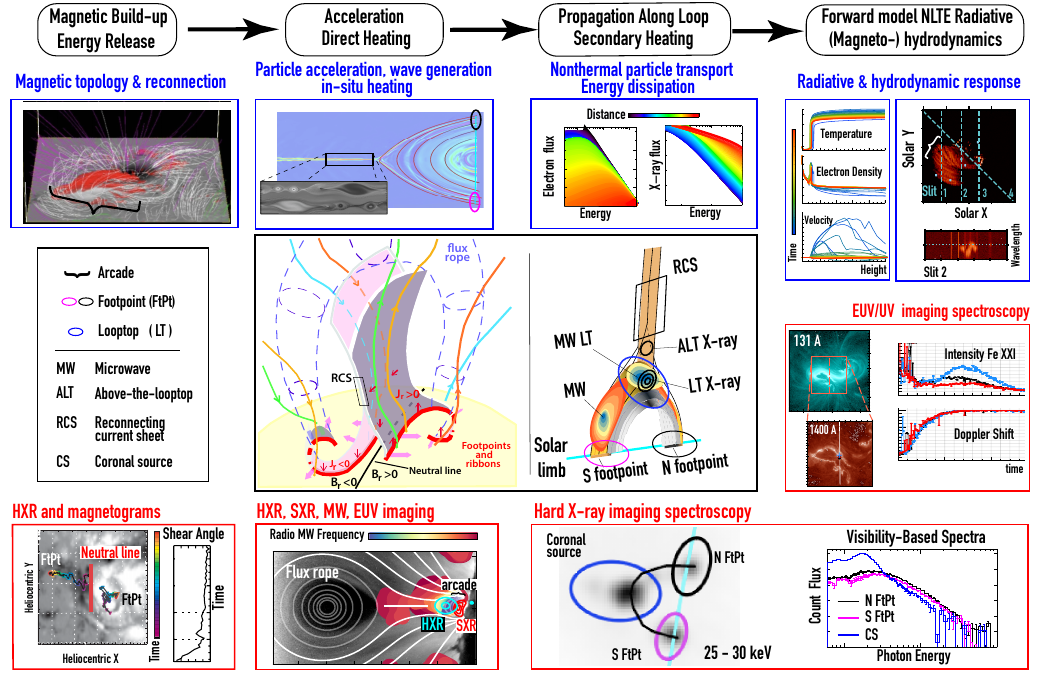}
\caption{\textsl{A pictorial schematic diagram demonstrating how component models are linked together into a SEE modeling framework and how these models can be tested by the observations. Starting from the top left corner: (\textbf{\#1}) 3D MHD simulations driven by footpoint shearing will be performed to follow the evolution of the magnetic field as energy builds and is rapidly released during reconnection. The magnetic configuration obtained from these models will be used as input to (\textbf{\#2}) particle acceleration models, which predict nonthermal electron and ion distributions and superhot thermal distributions. These will be compared with X-ray and microwave spectral imaging. The output nonthermal distributions from these will be input to (\textbf{\#3}) particle transport models, which track the transport of these high energy particles until their thermalization in the footpoints. Observed spectral images will be compared with the transport model predictions. Heating rates from the transport models will be used as input to (\textbf{\#4}) 1D loop models to predict the radiation and hydrodynamic response resulting from the heating produced by the accelerated particles as they propagate from the acceleration site to a footpoint. Using the \radynarcade\ technique, these 1D simulations will be ``wrapped'' onto closed loops in the magnetic field ``skeleton'' arcade obtained using the 3D MHD models (\textbf{\#1}). Synthetic images, line-of-sight velocities and line intensities will be computed from the models and compared to images and spectral observations (from e.g., SDO, Solar Orbiter, IRIS, MUSE, Hinode, and EOVSA). \label{fig:schematic}}}
\end{figure} 
\end{landscape}

\newpage
\vspace{-0.175in}
\section{Data-Driven SEE Modeling Framework}
\vspace{-.175in}
Dominant physical processes in SEEs have scale sizes ranging from the global (large scale magnetic field evolution) to the kinetic (particle acceleration and transport), which occurs on scales some ten orders of magnitude smaller. Currently, no single model can resolve the evolution of all relevant processes in 3D. For example, the high-resolution ($\Delta z  = 64$ km) simulation of flare energy build-up and impulsive release using the \muram\ 3D radiation-MHD code \citep{2019NatAs...3..160C} is still far from resolving the flaring transition region, which has scale size less than 1 km \citep{2022ApJ...931...60A}. Accurately resolving the transition region is critical for predicting chromospheric evaporation and the resultant coronal temperatures and densities \citep{2013ApJ...770...12B}. Moreover, MHD models make the assumption that fluids are thermal, but \emph{nonthermal} particles are ubiquitous in flares. In fact, they are a primary source of energy input to the lower solar atmosphere. Currently, only 1D field-aligned loop models (e.g., \radyn, \flarix, \hydrad) have the efficiency to be run at sufficient resolution ($\Delta z \sim 1$ m) to fully resolve the transition region and to track the kinetic scale physics of nonthermal particle interactions. But of course, 1D loop models lack realistic flare geometries and miss cross-field interactions. Although no single model is yet sufficient, \textbf{many models already exist that encompass key aspects of SEE evolution. We advocate coupling these, thereby building a comprehensive chain of models} that together simulate the spatio-temporal evolution of the resulting phenomena: accelerated particles (electrons and ions), heated plasma, radiation, and mass motions. In this section, we identify the key components necessary to build-up a SEE modeling framework. A pictorial schematic diagram demonstrating how these components are linked and are tested with observational constraints is shown in Figure~\ref{fig:schematic}. 

\begin{wrapfigure}{r}{0.525\textwidth}
        \vspace{-0.25in}
        \includegraphics[scale = 0.525, clip = true, trim = 0.5cm 0.35cm 1.25cm 0cm]{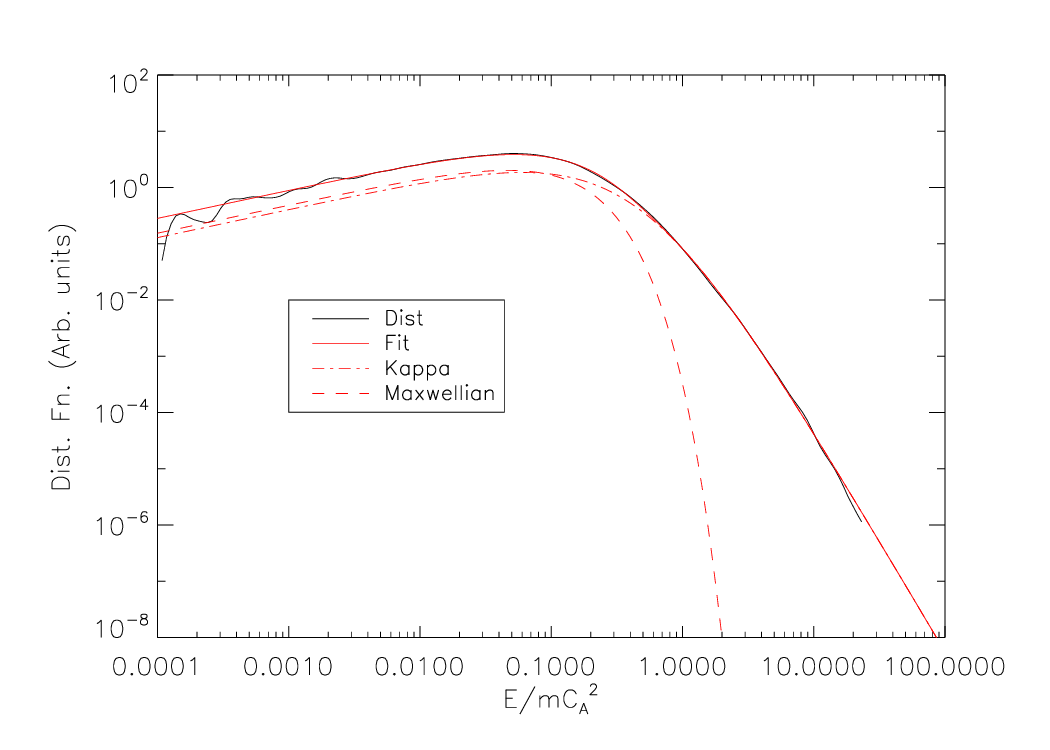}
        \vspace{-0.175in}
                \caption{{\textsl{Electron distribution function predicted by \kglobal. The x-axis is energy normalized by $m_I c_A^2$, which for typical flaring coronal conditions is approximately 10 keV. \kglobal\ well reproduces the two primary observables produced during flare acceleration, a power-law at higher-energy and a hot thermal component at lower-energy.}}\label{fig:kglobalspec}}
                \vspace{-.17in}
\end{wrapfigure}

\vspace{.1in}
\noindent \textbf{Component \#1: Modeling magnetic build-up and energy release}\\
The first component for comprehensive SEE modeling will be global-scale 3D MHD models, such as \arms\ \citep{2012ApJ...760...81K,2016ApJ...820...60G,1989BAPS...34.1287D,2019ApJ...879...96D} and \muram\ \citep{2017ApJ...834...10R} to track the evolution of the magnetic field and eventual reconnection in the flare current sheet (CS). The initial states can be obtained from coronal field extrapolations with boundary conditions from vector magnetograms \citep[e.g.,][and references therein]{2016NatCo...711522J,2022Innov...300236J}. These will provide crucial information on the evolving magnetic-field configuration, reconnection-associated flows, onset of energy deposition into the flare arcade, and burstiness/temporal variation of the magnetic energy release. In particular, these calculations must track the magnetic connectivity between the proposed origins of the accelerated particles (e.g., plasmoids) and  their destinations (e.g., the flare ribbons). Key properties such as the magnetic shear (guide field) on the newly reconnected field lines can be measured throughout the eruptive flare. The magnetic configuration at reconnection onset, amount of shear  and characteristics of plasmoid can be used to initiate simulations of the flare particle acceleration process (\textbf{\#2}). The changing flare-arcade geometry and energy injection identified in the MHD simulations will serve as input for an ensemble of 1D loops models (\textbf{\#4}) to derive flare emissions. 

\vspace{.1in}
\noindent \textbf{Component \#2: Modeling particle acceleration and direct heating}\\
\vspace{-0.2in}

The dominant mechanism for flare particle acceleration remains an open question. Stochastic acceleration by turbulence \citep{2012SSRv..173..535P} and Fermi reflection in contracting and merging magnetic flux ropes \citep{2006Natur.443..553D, 2014PhPl...21i2304D} are efficient accelerators of nonthermal electrons and ions \citep{2017ApJ...843...21L}. The configurations of magnetic fields including plasmoid properties produced from \textbf{\#1} should be used as initial states for models of flare particle acceleration. 

\begin{wrapfigure}{r}{0.6\textwidth}
\vspace{-0.2in}
\includegraphics[scale = 0.4, clip = true, trim = 0.cm 0.cm 0cm 0cm]{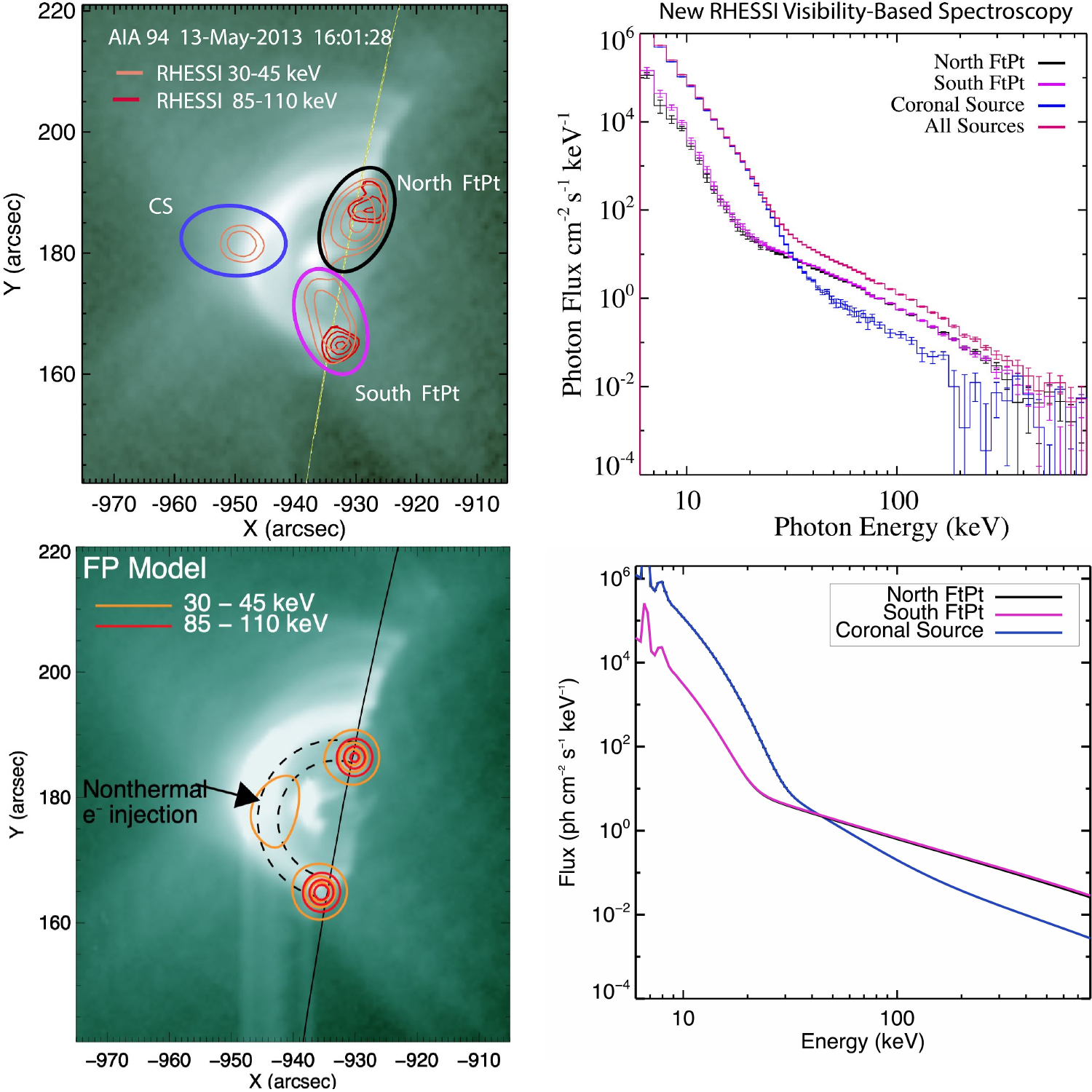}
\vspace{-0.175in}
\caption{\textsl{(Top left panel) AIA 94 image during the SOL2013-05-13T16:01 (X2.8) solar flare with RHESSI HXR contours overlaid. (Top right panel) HXR spectra of the footpoints and a coronal source. Nonthermal emission (\textgreater{} 30 keV) is clearly evident in the coronal source. (Bottom panels) \fp\ model of the locations (left panel) and spectra (right panel) of HXR emission in this flare. Nonthermal electrons are injected at the top of the model coronal loop (marked by dashed lines) which has a strongly converging magnetic field, trapping particles near the top and producing a nonthermal coronal source.}\label{fig:rhessispec}}
\vspace{-.3in}
\end{wrapfigure}

These then predict the directly heated thermal as well as the nonthermal (typically, a power-law) components of the accelerated particles. As an example, we show in Figure~\ref{fig:kglobalspec} a prediction of the \kglobal\ model \citep{2019PhPl...26a2901D, 2021PhRvL.126m5101A} that produces both hot thermal and nonthermal (power-law) electrons during flare energy release. The directly heated component, detected by thermal X-ray bremsstrahlung, can be compared to observations \citep[e.g.,][]{2014ApJ...781...43C}. But the nonthermal particles move quickly through the legs of flares loops interacting with and heating the ambient plasma along their transport. Comparing predicted nonthermal bremsstrahlung and gyrosynchrotron emission resulting from these interactions with observations requires particle transport models (\textbf{\#3}).

\vspace{0.1in} 
\noindent \textbf{Component \#3: Modeling nonthermal particle transport}\\
Predicted nonthermal distributions from \textbf{\#2} will be injected into models of particle transport \citep[e.g., \fp;][]{2020ApJ...902...16A}, which track nonthermal particle distributions as they travel down the flux tubes to the footpoints. Electrons emit detectable gyrosynchrotron microwave emission along the legs and, many megameters below, bremsstrahlung X-rays as they impact the denser plasma in the chromosphere. During their transport, the nonthermal electron distribution evolves under the influence of many forces, especially the return-current electric field \citep{2012ApJ...745...52H, 2017ApJ...851...78A, 2021ApJ...917...74A}. Until recently, the evolution of these particles during their transport has been poorly modeled, preventing the direct comparison of X-ray observations in the flare footpoints with acceleration model predictions. Computational models such as \fp\ solve the Fokker-Planck equation accounting for energy loss and pitch-angle diffusion from the dominant forces including  Coulomb collisions and the return-current electric field and potentially including the presence of runaway electrons \citep{2021ApJ...917...74A}. An important feature of these transport models is their ability to predict particle distributions at any position along flux tubes --not just in the thick-target footpoints-- allowing a direct comparison to HXR and microwave imaging spectroscopy observations.

An example is shown in Figure~\ref{fig:rhessispec}. Two compact footpoint HXR sources near the solar limb in this event are clearly seen in the top left panel. There is also a coronal nonthermal source but it is barely evident, because it is more extended and the flux per pixel is very low. Spectra for the two footpoints and coronal source are  in the top right-hand panel. The nonthermal emission, evident at the looptop, means that accelerated electrons are trapped and lose energy high in the loop. The bottom panels of Figure~\ref{fig:rhessispec} show images and spectra using the \fp\ model with electrons injected into a flux tube with strongly converging magnetic field, thereby trapping nonthermal electrons. Spatially-resolved transport models such as \fp\ are precisely the right tools to investigate the cause of this trapping, and to determine the electron distribution at the acceleration region thereby constraining acceleration models (\textbf{\#2}).

\begin{wrapfigure}{r}{0.7\textwidth}
\vspace{-0.175in}
\includegraphics[scale = 0.4, clip = true, trim = 0.cm 1.5cm 0cm 1.5cm]{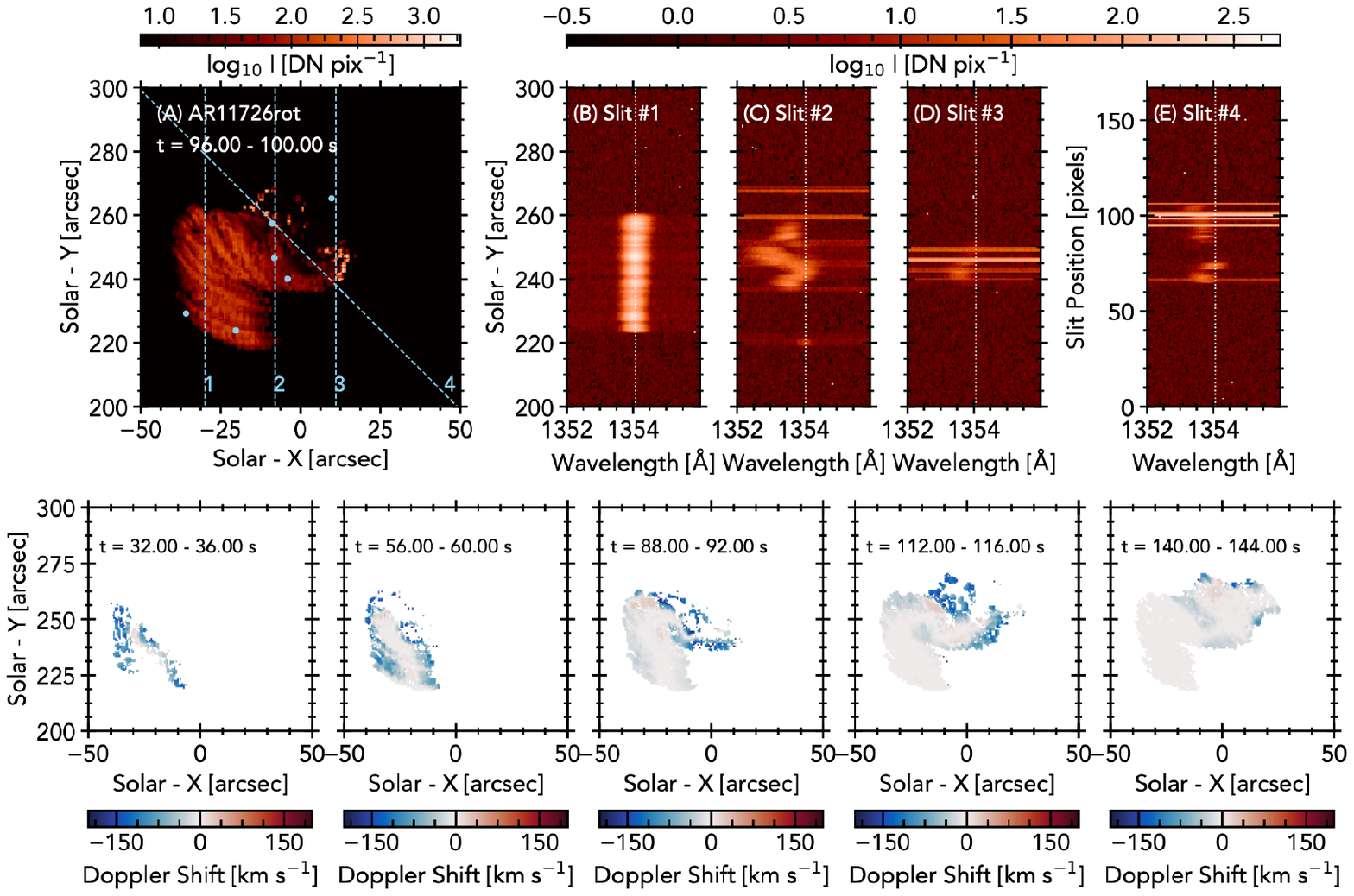}
\vspace{-0.15in}
\caption{\textsl{A \radynarcade\ simulation of a flare arcade \citep{2020ApJ...900...18K} Panel (A) shows an image of the flare arcade observed in the 1352--1356~\AA\ wavelength range. Slit positions are indicated. Panels (B-E) show the \ion{Fe}{xxi} spectral line, where synthetic spectra have been folded through the IRIS instrumental response. The lower panels show the line-of-sight Doppler shifts of the \ion{Fe}{xxi} line obtained from Gaussian fitting. The greyscale in those images is the line intensity to illustrate that Doppler shifts are strongest at the base of the modeled loops.}\label{fig:radynarcade}}
\vspace{-.275in}
\end{wrapfigure}

As the injected nonthermal particles collide with the ambient plasma, they heat it, driving fast flows, elevating densities, and producing intense bursts of radiation. Heating rates predicted by particle transport models will then be used to model the radiative hydrodynamic response of flare loops (\textbf{\#4}).

\vspace{.1in}
\noindent \textbf{Component \#4: Modeling the  radiation hydrodynamic evolution of flare loops}\\
The heating rates predicted from \textbf{\#3} will be input to state-of-the-art 1D loop models such as \radyn\ \citep{1992ApJ...397L..59C, 1995ApJ...440L..29C, 1999ApJ...521..906A, 2005ApJ...630..573A, 2015ApJ...809..104A}, \hydrad\ \citep{2003A&A...407.1127B, 2013ApJ...770...12B}, and \flarix\ \citep{2016IAUS..320..233H}. These models track the radiative hydrodynamic response to flare heating on the intrinsic temporal ($10^{-5}$~s) and spatial (1~m) scales of the transition region and chromosphere. Importantly, they are able to model the optically-thick non-LTE radiative transfer and non-equilibrium atomic level populations that dominate the chromosphere where much of the flare energy is deposited.  Since they resolve the transition region, they are able to accurately predict the responding chromospheric evaporation flows and coronal densities and temperatures. 

Since loop models run efficiently, numerous individual loops can be modeled. Loop lengths and orientations will be obtained from field line tracing in the MHD models (\textbf{\#1}). Then the predicted temperatures, densities and flow velocities can be wrapped back onto these field lines, thus populating the 3D volume with a realistic model of a flare arcade. In this way, comprehensive and data-driven 3D models of SEEs on their intrinsic spatial and temporal scales can be constructed. An example of this, using the \radynarcade\ technique \citep{2020ApJ...900...18K} is shown in Figure~\ref{fig:radynarcade}. \ion{Fe}{xxi} line emission was forward modeled, and degraded to IRIS spectral and spatial resolution. The line profiles exhibit shifts, asymmetries, broadenings and intensity enhancements. The resulting Doppler shifts are comparable to those observed, and are strongest at the feet and lower legs of the loops, consistent with observations.   

\vspace{-0.175in}
\section{Comparing Resulting Emissions with Observations}
\vspace{-0.15in}
The end result of the SEE modeling framework will be a data-driven time-dependent model of SEE initiation and the resulting flare arcade. From this, emissions from numerous mechanisms and passbands can be constructed and compared directly to observations. For example, thermal soft X-ray bremsstrahlung, predicted using the modeled temperatures and emission measures of hot plasma, can be passed through GOES responses and then compared directly with GOES light curves. Even more, wavelength-dependent emissivities of numerous atomic transitions from CHIANTI \citep{1997A&AS..125..149D, 2021ApJ...909...38D} can be modeled within the 3D volume. These will be projected along an observer's line of sight, and passed through instrumental responses to construct synthetic 2D images that can be compared directly to observations, for example from SDO/AIA, Solar Orbiter/EUI and IRIS/Slit Jaw Imager. Synthetic spectral images will also be produced using emissivities with Doppler shifts obtained by projecting the bulk flow along the line-of-sight. These will be passed through responses of imaging spectrographs such as EUVST, Solar Orbiter/SPICE, MUSE, IRIS, and Hinode/EIS and compared directly to these instruments. \textbf{Since the framework is a set of chained models, observations of one piece of the chain can be used to constrain models in different pieces.} For example, hard to observe processes, such as flare acceleration (\textbf{\#2}) can be constrained by the magnetic field configuration (\textbf{\#1}), from the observed nonthermal bremsstrahlung (predicted by \textbf{\#3}), and the resulting heating and emissions (\textbf{\#4}).     

\vspace{-0.175in}
\section{Recommendations}
\vspace{-0.15in}
The computational modeling framework that we have outlined will require significant investment to bring together a large body of researchers to coordinate developing and integrating new models and comparing them with observations. Just as large resources are allocated for building next-generation instrumentation, so too, sufficient funding must be dedicated to building next-generation models. Model and instrument development naturally go together, since models inform instrument builders on which observations will be most discriminating for testing and refining theories. Our framework will ensure that future SEE-observing missions (e.g., see white papers by Caspi et al. on the COMPLETE, Christe et al. on the FOXSI, Shih et al. on the FIERCE, and Gary et al. on the FASR mission concepts) are informed by state-of-the-art models. 

Current theory and modeling funding opportunities are typically relatively small and are not sufficient to fully realize comprehensive SEE models. Therefore, \textbf{we recommend  large-scale increased investment in theory and modeling with the goal of producing comprehensive SEE models.} Possible mechanisms are through dedicated DRIVE Centers or the creation of new joint NSF/NASA funding opportunities. Understanding the energy release processes in SEEs and the related particle-acceleration processes are major goals in Heliophysics, with significant societal benefits from better understanding and prediction of destructive space weather at its source.
\newpage
\pagenumbering{gobble}
\bibliography{dswp.bib}
\end{document}